\documentclass[twocolumn]{aastex631}
\usepackage{latexsym}
\usepackage{graphicx}
\usepackage{color}
\usepackage{amsmath}

\usepackage{hyperref}
\usepackage{bm}
\usepackage{acronym}
\usepackage{url}
\usepackage{textcomp}
\usepackage{xspace}
\usepackage{acronym}
\usepackage{multirow}

\newcommand{\scinum}[2]{\ensuremath{#1\!\times\!10^{#2}}}
\newcommand{\figref}[1]{\figurename~\ref{#1}}
\newcommand{\tabref}[1]{\tablename~\ref{#1}}
\newcommand{\secref}[1]{\ref{#1}}
\newcommand{\code}[1]{\texttt{#1}\xspace}

\begin{document}

\reportnum{}

\title{Searches for Gravitational Waves from Known Pulsars at Two Harmonics in the Second and Third LIGO-Virgo Observing Runs}

\author{R.~Abbott}
\affiliation{LIGO Laboratory, California Institute of Technology, Pasadena, CA 91125, USA}
\author{H.~Abe}
\affiliation{Graduate School of Science, Tokyo Institute of Technology, Meguro-ku, Tokyo 152-8551, Japan  }
\author{F.~Acernese}
\affiliation{Dipartimento di Farmacia, Universit\`a di Salerno, I-84084 Fisciano, Salerno, Italy  }
\affiliation{INFN, Sezione di Napoli, Complesso Universitario di Monte S. Angelo, I-80126 Napoli, Italy  }
\author[0000-0002-8648-0767]{K.~Ackley}
\affiliation{OzGrav, School of Physics \& Astronomy, Monash University, Clayton 3800, Victoria, Australia}
\author[0000-0002-4559-8427]{N.~Adhikari}
\affiliation{University of Wisconsin-Milwaukee, Milwaukee, WI 53201, USA}
\author[0000-0002-5731-5076]{R.~X.~Adhikari}
\affiliation{LIGO Laboratory, California Institute of Technology, Pasadena, CA 91125, USA}
\author{V.~K.~Adkins}
\affiliation{Louisiana State University, Baton Rouge, LA 70803, USA}
\author{V.~B.~Adya}
\affiliation{OzGrav, Australian National University, Canberra, Australian Capital Territory 0200, Australia}
\author{C.~Affeldt}
\affiliation{Max Planck Institute for Gravitational Physics (Albert Einstein Institute), D-30167 Hannover, Germany}
\affiliation{Leibniz Universit\"at Hannover, D-30167 Hannover, Germany}
\author{D.~Agarwal}
\affiliation{Inter-University Centre for Astronomy and Astrophysics, Pune 411007, India}
% ...

\collaboration{0}{The LIGO Scientific Collaboration, the Virgo Collaboration, and the KAGRA Collaboration}

\date{\today}

\begin{abstract}         
\noindent
We present a targeted search for continuous gravitational waves (GWs) from 236 pulsars using data from the third observing run of LIGO and Virgo (O3) combined with data from the second observing run (O2). Searches were for emission from the $l=m=2$ mass quadrupole mode with a frequency at only twice the pulsar rotation frequency (single harmonic) and the $l=2, m=1,2$ modes with a frequency of both once and twice the rotation frequency (dual harmonic). No evidence of GWs was found so we present 95\% credible upper limits on the strain amplitudes $h_0$ for the single harmonic search along with limits on the pulsars' mass quadrupole moments $Q_{22}$ and ellipticities $\varepsilon$. Of the pulsars studied, 23 have strain amplitudes that are lower than the limits calculated from their electromagnetically measured spin-down rates. These pulsars include the millisecond pulsars J0437\textminus4715 and J0711\textminus6830 which have spin-down ratios of 0.87 and 0.57 respectively. For nine pulsars, their spin-down limits have been surpassed for the first time. For the Crab and Vela pulsars our limits are factors of $\sim 100$ and $\sim 20$ more constraining than their spin-down limits, respectively. For the dual harmonic searches, new limits are placed on the strain amplitudes $C_{21}$ and $C_{22}$. For 23 pulsars we also present limits on the emission amplitude assuming dipole radiation as predicted by Brans-Dicke theory.
\end{abstract}

\section{Introduction}
\label{sec:introduction}

To date, the LIGO and Virgo observatories have made detections of numerous sources of gravitational radiation. These detections have been of transient gravitational waves (GWs) from the inspiral and subsequent mergers of compact binary objects including binary black holes and binary neutron stars \citep{abbott2021gwtc2}. Recently, the list of observed events expanded to include neutron star-black hole binaries \citep{2021ApJ...915L...5A}. There remain other types of GW sources that are yet to be observed such as continuous GW (CW) sources. Unlike transients, CW signals are almost monochromatic, with their amplitude and frequency changing very slowly over year-long timescales. The mass quadrupoles of these sources, such as deformed neutron stars, are expected to be far smaller than those involved in compact binaries and therefore only local galactic sources are likely to produce detectable signals.

Likely candidates for producing such signals are neutron stars spinning with some non-axisymmetric deformation \citep{1979PhRvD..20..351Z}. This may be a solid deformation such as mountains on the crust produced during cooling \citep{2000MNRAS.319..902U}, from binary accretion \citep{gittins2021modelling} or due to strong magnetic fields \citep{1996AA...312..675B, 2002PhRvD..66h4025C}.
GW radiation can also be caused by fluid modes of oscillation beneath the crust such as r-modes \citep{1998ApJ...502..708A, 1998ApJ...502..714F}.
By detecting CWs, light can be shed on the structure of the star. Additionally, detections of such GWs can be used to test general relativity via the constraint of non-standard GW polarization \citep{2017PhRvD..96d2001I, PhysRevD.100.104036}.
A more thorough discussion of various methods of GW emission from neutron stars can be found in \citet{2017MPLA...3230035R, 2018ASSL..457..673G}.

The structure of this paper is as follows. Section~\secref{sec:search_types} outlines the types of CW searches. Section~\secref{sec:sig_models} describes the types of signal models used in this analysis. Section~\secref{sec:analysis} describes the search methods used. Section~\secref{sec:data} covers both the GW and EM data used. We present our results in Section~\secref{sec:results} with conclusions in Section~\secref{sec:conclusions}.

\subsection{Continuous-wave searches}
\label{sec:search_types}

There are broadly three types of CW searches. Targeted searches look for signals from known pulsars for which their rotational phases can be accurately determined from electromagnetic (EM) observations \citep[e.g.,][]{2017ApJ...839...12A, 2019ApJ...879...10A, Nieder19, Nieder20, Abbott_2020, 2021ApJ...913L..27A, 2021ApJ...923...85A}. This simplifies the search as the EM observations can be used to derive a timing solution and it is assumed that the GW phase evolution is locked to the EM evolution. This means the search is over a small parameter space, generally limited to the unknown signal amplitude and orientation of the source, which allows a more sensitive search than other methods. In some targeted searches, the assumption that the GW evolution follows the EM evolution is relaxed and the search is performed in a narrow band around the expected frequency and spin-down rate \citep{2017PhRvD..96l2006A, O2Narrowband}. In this case, the search is more computationally expensive due to the larger parameter range being searched and slightly less sensitive because of a higher trials factor. To overcome this, narrow-band searches often look at fewer targets.
Directed searches look for signals from small sky regions that are believed to have a high probability of containing a neutron star, such as supernova remnants. As the timing solution cannot be derived from EM observations, a wide range of rotational parameters must be searched.
All-sky searches look for signals in all sky directions and over a wide range of rotational parameters. A review of directed and all-sky methods and previous searches can be found in \citet{Tenorio:2021wmz}. Both these methods suffer increasing computational costs and decreasing sensitivity of the searches as parameter space increases.

Searches of all three types have been performed and so far no convincing evidence for CWs has been observed. However, searches have probed new regimes, such as providing upper limits on emission that are more stringent (i.e., smaller) than those based on energetics arguments. For example, for several pulsars including the Crab pulsar, Vela pulsar \citep{2019ApJ...879...10A}, J0537\textminus6910 \citep{2021ApJ...913L..27A, 2021arXiv210414417T} and two millisecond pulsars \citep{Abbott_2020} the direct upper limits set on the GW amplitude are more constraining than limits based on the assumption that all the pulsars' spin-down luminosity is radiated through GWs, known as the spin-down limit.

In this paper we report the new results of a targeted search for CW signals from 236 pulsars using the most recent LIGO and Virgo datasets including the second and third observing runs (O2 and O3 respectively). LIGO \citep{2015CQGra..32g4001L} consists of two gravitational-wave detectors situated in Hanford, Washington (H1) and Livingston, Louisiana (L1) while Virgo \citep{2015CQGra..32b4001A} is located in Cascina, Pisa (V1). 
The ephemerides for the pulsars have been derived from observations using the CHIME, Hobart, Jodrell Bank, MeerKAT, Nancay, NICER and UTMOST observatories. More details on these observations can be found in Section~\secref{sec:data_em}.

\subsection{Signal models}
\label{sec:sig_models}

We assume that the gravitational-wave (GW) emission is locked to the rotational phase of the pulsar. For the ideal case of a triaxial star rotating steadily about a principal moment of inertia axis, the GW emission is at twice the star's spin frequency, $f_{\rm rot}$.
However, there are mechanisms that can produce variations to this $2f_{\rm rot}$ frequency. For example, a superfluid component with a misaligned spin axis within the star could produce a dual-harmonic emission at both once and twice the rotation frequency without leaving an imprint on the EM emission \citep{2010MNRAS.402.2503J}. Therefore we perform two searches: one at just twice the pulsar rotation frequency and one at both one and two times the frequency, which is referred to as a dual harmonic search.

The waveform used in the dual harmonic search is detailed in \citet{2010MNRAS.402.2503J} and used in \citet{2015MNRAS.453.4399P, 2017ApJ...839...12A, 2019ApJ...879...10A, Abbott_2020}. The signals $h_{21}$ and $h_{22}$ at once and twice the pulsar rotation frequency can be defined as
\begin{eqnarray}
\nonumber
\label{eq:h21}
h_{21} &= - \frac{C_{21}}{2} \Big[&F^D_+(\alpha, \delta, \psi; t) \sin\iota \cos\iota \cos{\left(\Phi(t)+ \Phi_{21}^C\right)} +  \\
& & F^D_\times(\alpha, \delta, \psi; t)\sin\iota  \sin{\left(\Phi(t)+ \Phi_{21}^C\right)} \Big] ,  \\
\nonumber
\label{eq:h22}
h_{22} &= - C_{22} \Big[&F^D_+(\alpha, \delta, \psi; t) (1 + \cos^2\iota) \cos{\left(2\Phi(t)+ \Phi_{22}^C\right)} +   \\
& & 2 F^D_\times(\alpha, \delta, \psi; t) \cos\iota  \sin{\left(2\Phi(t)+ \Phi_{22}^C\right)} \Big] ,
\end{eqnarray}
where $C_{21}$ and $C_{22}$ are the dimensionless constants that give the component amplitudes, the angles $(\alpha, \delta)$ are the right ascension and declination of the source, while the angles $(\iota, \psi)$ describe the orientation of the source's spin axis with respect to the observer in terms of inclination and polarization, $\Phi_{21}^C$ and $\Phi_{22}^C$ are phase angles at a defined epoch and $\Phi(t)$ is the rotational phase of the source. The antenna functions $F^D_+$ and $F^D_\times$ describe how the two polarization components (plus and cross) are projected onto the detector. 

For the ideal case of a steadily spinning triaxial star emitting GWs only at twice the rotation frequency, the equatorial ellipticity can be defined as
\begin{equation}
    \varepsilon \equiv \frac{|I_{xx} - I_{yy}|}{I_{zz}} ,
\end{equation}
where $(I_{xx}, I_{yy}, I_{zz})$ are the source's principal moments of inertia, with the star rotating about the $z$-axis. The mass quadrupole of the source $Q_{22}$ is often quoted and is related to the ellipticity as
\begin{equation}
    Q_{22} = I_{zz} \varepsilon \sqrt{\frac{15}{8\pi}} .
    \label{eq:q22}
\end{equation}
For single harmonic emission, $C_{21}$ from equation \eqref{eq:h21} can be set as 0, leaving only $C_{22}$ in equation \eqref{eq:h22}. The amplitude can then be parameterised as the dimensionless $h_0$: the amplitude of the circularly polarised signal that would be observed if the source lay directly above or below the plane of the detector and had its spin axis pointed directly towards or away from the detector. The following equations are defined in \citet{2014ApJ...785..119A}.
\begin{equation}\label{eq:h0}
    h_0 = 2C_{22} = \frac{16\pi^2 G}{c^4} \frac{I_{zz} \varepsilon f_{\rm rot}^2 }{d} ,
\end{equation}
where $d$ is the distance of the source.
The spin-down limit $h_0^{\rm sd}$ of a source is given by:
\begin{equation}\label{eq:spin-down}
    h_0^{\rm sd}=\frac{1}{d}\left(\frac{5GI_{zz}}{2c^3}\frac{|\dot{f}_{\rm rot}|}{f_{\rm rot}}\right)^{1/2} ,
\end{equation}
where $\dot{f}_{\rm rot}$ is the first derivative of the rotational frequency, i.e., the spin-down rate,
and provides an amplitude limit assuming that all of the rotational energy lost by the pulsar is converted to gravitational-wave energy \citep{2005PhRvL..95u1101O}. When the directly observed $h_0$ is smaller than $h_0^{\rm sd}$, the spin-down limit can be said to have been surpassed. This information is most often represented by quoting the ``spin-down ratio'', i.e., the ratio between $h_0$ and $h_0^{\rm sd}$. If assuming that there is no mechanism (e.g., accretion) providing some additional spin-up torque, the direct amplitude constraints probe a new physical regime only when the spin-down limit is surpassed.
There are two types of spin-down rate: intrinsic and observed. The observed spin-down rate can be affected by the transverse velocity of the source, e.g., the Shlovskii effect \citep{1970SvA....13..562S}, so where possible the intrinsic spin-down rate is used to calculate the spin-down limit.

\subsubsection{Non-GR polarization signal}
\label{sec:sig_models_BD}

In this paper we also perform a search for GWs with polarizations as predicted in a modification of the standard general relativity (GR) proposed by Brans and Dicke. The Brans-Dicke theory \citep{1961PhRv..124..925B} predicts three independent polarization states: two tensor polarizations, as in GR, and an additional scalar polarization. The dominant scalar radiation in Brans-Dicke theory originates from the time-dependent dipole moment \citep[see][for details]{universe7070235}. The signal $h_{11}$ due to dipole radiation is given by \citep[see Eqs.~(63) - (67) and (70) - (71) of][]{universe7070235} 
\begin{equation}\label{eq:h11}
    h_{11} = - h_{0}^d \, c(\alpha,\delta;t) \, \sin \iota \sin (\Phi(t) + \Phi_0),
\end{equation}
where $c(\alpha,\delta;t)$ is the amplitude modulation function and $\Phi_0$ is the phase angle at time $t = 0$. The explicit formula for $c(\alpha,\delta;t)$ is given by Eq.~(64) of \citet{universe7070235}. We see that the dipole radiation comes at the rotational frequency of the pulsar.
We assume that the only non-vanishing component $D$ of the dipole moment in the pulsar's frame is in the $x$-direction. Then the amplitude $h_0^d$ of the signal is given by 
\begin{equation}\label{eq:hoS}
    h_{0}^d = \frac{4 \pi G}{c^3} \zeta \frac{D f_{\rm rot}}{d},
\end{equation}
where $\zeta$ is the parameter of the Brans-Dicke theory \citep[see Section~3 of ][for details]{universe7070235}.

\section{Analysis}
\label{sec:analysis}

As in \citet{2017ApJ...839...12A, 2019ApJ...879...10A}, three largely independent analysis methods were used for the searches in this paper: the \textit{Time-domain Bayesian method} (Section~\secref{sec:bayesian}), the $\mathcal{F}$/$\mathcal{G}$/$\mathcal{D}$-statistic method (Section~\secref{sec:fstat}), and the \textit{5n-vector method} (Section~\secref{sec:5vect}). The $\mathcal{F}$/$\mathcal{G}$/$\mathcal{D}$-statistic and the 5n-vector methods are only used in searches for pulsars deemed to be high-value: those which surpass their spin-down limits in the Bayesian analysis. The Bayesian and $\mathcal{F}$/$\mathcal{G}$/$\mathcal{D}$-statistic methods search for two signal models: a single harmonic signal emitted by the $l=m=2$ mass quadrupole mode at twice the pulsar rotation frequency and a dual harmonic signal emitted by the $l=m=2$ {\it and} $l =2, m=1$ modes at twice and once the frequency. The 5n-vector method restricts the latter search to the $l =2, m=1$ mode only. Only one method, the 
$\mathcal{D}$-statistic, is used for the Brans-Dicke polarization search. The GW emission is assumed to precisely follow the phase evolution determined through EM observations, although uncertainties in values from the EM observations are not accounted for here.

\subsection{Time-domain Bayesian method}
\label{sec:bayesian}

The raw GW strain data are heterodyned using their expected phase evolution, which includes corrections for the relative motion of the source with respect to the detector and relativistic effects \citep{Dupuis:2005}. They are then low-pass filtered using a cut-off frequency of $0.25$\,Hz and then down-sampled to one sample per minute (1/60\,Hz bandwidth) centred about the expected signal frequency now at 0\,Hz. For the dual harmonic search this is repeated so that a time series is obtained centred at both $f_{\rm rot}$ and $2f_{\rm rot}$. Bayesian inference is used to estimate the remaining unknown signal parameters and the evidence for the signal model \citep{2017arXiv170508978P}. For the parameter inference, the prior distributions used were those given in Appendix~2 of \citet{2017ApJ...839...12A}. They were uninformative uniform priors for the orientation angles, unless restricted ranges were appropriate as discussed in Section~\secref{sec:restricted}. For the amplitude parameters, Fermi-Dirac distribution priors were used \citep[see Section 2.3.5 of][]{2017arXiv170508978P}. The Fermi-Dirac distributions for each pulsar were set such that they were close to flat over the bulk of the likelihood while penalizing very large values. This choice of prior means that the amplitude posteriors will be dominated by the likelihood even when no signal is observed. Any upper limits derived from the posteriors will be more conservative than those that would be found from using an uninformative Jeffreys prior that was uniform in the logarithm of the amplitude, i.e., $p(h_0) \propto 1 / h_0$. To avoid basing the priors on current detector data the priors were constructed by choosing Fermi-Dirac parameters that gave distributions for which the 95\% probability upper bound was equivalent to the estimated upper limit sensitivity of the combined LIGO S6 and Virgo VSR4 science runs at the particular pulsar GW signal frequency.\footnote{For the dual harmonic search for pulsars with signal components below 20\,Hz, the $C_{21}$ priors were constructed without using the estimated VSR4 Virgo sensitivity. This was to prevent prior from dominating over the likelihood in this frequency region.}

This method also considers the effect of glitches on the pulsars (Section~\secref{sec:glitches}) and can perform searches with restrictions on the pulsar orientation (Section~\secref{sec:restricted}).

\subsubsection{Glitches}
\label{sec:glitches}

Although their frequency is usually very stable, pulsars occasionally experience a transient increase in rotation frequency and frequency derivative. Such events are called glitches and are most common in young, non-recycled pulsars, although they do rarely occur in millisecond pulsars \citep{2004ApJ...612L.125C, 2016MNRAS.461.2809M}. Some of our sample of pulsars glitched during the course of O2 and O3. We assume that glitches affect the GW phase identically to the EM phase, but with the addition of an unknown phase offset at the time of the glitch. This phase offset is included in the parameter inference. For glitches that occur before or after the range of the data, no phase offset is needed. The pulsars which experienced glitching during the course of this analysis are J0534+2200, also known as the Crab pulsar \citep{2021MNRAS.505L...6S}, J0908\textminus4913 \citep{2019RNAAS...3..192L} and J1105\textminus6107. They are shown in \tabref{tab:glitches} along with the time (MJD) of the glitch.

\begin{table*}[htp]
    \caption{Pulsars with glitches occurring during the course of the runs used in this analysis.} 
    \centering
    \begin{tabular}{|c|c|}
        \hline
        PSR & Epoch (MJD)\\
        \hline
        J0534+2200 (Crab) & 58687.6448 $\pm$ 0.0033\\
        J0908\textminus4913 & 58767.34 $\pm$ 4.5 \\
        J1105\textminus6107 & 58582.24 \\
        \hline
    \end{tabular}
    \label{tab:glitches}
\end{table*}
\subsubsection{Restricted orientations}
\label{sec:restricted}

Occasionally, the orientation of a pulsar can be determined from modeling of X-ray observations of its pulsar wind nebula \citep{Ng:2004,Ng:2008}. In such cases, these values can be included as narrow priors on inclination and polarization angle rather than using an uninformative uniform prior. Results still assuming uniform priors are also included. Such pulsars are shown in \tabref{tab:restricted_priors} below along with their restricted prior ranges \citep{2017ApJ...839...12A}, which are assumed to be Gaussian about the given mean and standard deviation.

\begin{table*}
    \centering
    \caption{Pulsars with observations sufficient to restrict their orientation priors in terms of inclination and polarization and the values used as the constraints.}
    \begin{tabular}{|c|c|c|c|}
        \hline
        PSR & $\Psi$ (rad) & $\iota_1$ (rad) & $\iota_2$ (rad)\\
        \hline
        J0534+2200 (Crab) & 2.1844 $\pm$ 0.0016 & 1.0850 $\pm$ 0.0149 & 2.0566 $\pm$ 0.0149 \\
        J0835-4510 (Vela) & 2.2799 $\pm$ 0.0015 & 1.1048 $\pm$ 0.0105 & 2.0368 $\pm$ 0.0105 \\
        J1952+3252 & 0.2007 $\pm$ 0.1501 & & \\
        J2229+6114 & 1.7977 $\pm$ 0.0454 & 0.8029 $\pm$ 0.1100 & 2.3387 $\pm$ 0.1100 \\
        \hline
    \end{tabular}
    \label{tab:restricted_priors}
\end{table*}

In the case of the Crab pulsar, which both experienced a glitch and has sufficient observations for restricted priors, four individual analyses are performed. Each analysis accounts for the glitch, with combinations of dual/single harmonic search and restricted/unrestricted priors. 

\subsection{\texorpdfstring{$\mathcal{F}$}{\textit{F}}/\texorpdfstring{$\mathcal{G}$}{\textit{G}}/\texorpdfstring{$\mathcal{D}$}{\textit{D}}-statistic method}
\label{sec:fstat}
The time-domain $\mathcal{F}$/$\mathcal{G}$/$\mathcal{D}$-statistic method uses the $\mathcal{F}$-statistic derived in \citet{1998PhRvD..58f3001J} and the $\mathcal{G}$ statistic derived in \citet{Jaranowski:2010}. The input data for the analysis using this method are the heterodyned data used in the time-domain Bayesian analysis.
The $\mathcal{F}$-statistic is used when the amplitude, phase and polarization of the signal are unknown, whereas the $\mathcal{G}$-statistic is applied when only amplitude and phase are unknown, and the polarization of the signal is known (as described in Section~\ref{sec:restricted}). The methods have been used in several analyses of LIGO and Virgo data \citep{2011ApJ...737...93A, 2014ApJ...785..119A, 2017ApJ...839...12A,Abbott_2020}.
The method also uses the $\mathcal{D}$-statistic developed in \citet{universe7070235} to search for dipole radiation in Brans-Dicke theory. The $\mathcal{D}$-statistic search is the first search of LIGO and Virgo data for dipole radiation as predicted by Brans-Dicke theory.

In this method a signal is detected in the data
if the value of the $\mathcal{F}$-, $\mathcal{G}$- or $\mathcal{D}$-statistic exceeds a certain threshold corresponding to an acceptable false alarm probability. We consider the false alarm probability of 1\% for the signal to be significant. 
The $\mathcal{F}$-, $\mathcal{G}$-, and $\mathcal{D}$-statistics are computed for each detector and each inter-glitch period separately. The results from different detectors or  different inter-glitch periods are then combined incoherently by adding the respective statistics.
When the values of the statistics are not statistically
significant, we set upper limits on the amplitude of the
gravitational-wave signal.

\subsection{The 5$n$-vector method}
\label{sec:5vect}
The 5$n$-vector method, derived in \citet{Astone:2012}, is a multi-detector matched filter in the frequency domain, based on the sidereal modulation of the expected signal amplitude and phase. 
The method has been used in several analysis of LIGO and Virgo detector data \citep{2011ApJ...737...93A, 2014ApJ...785..119A, 2017ApJ...839...12A,Abbott_2020}, and recently \citet{abbott2021gwtc2} combined with the Band-Sample Data (BSD) framework \citep{bsd:2018}. This is based on the construction of BSD files, i.e., complex time series that cover $10$\,Hz and $1$ month of the original dataset. Using the BSD files, the computational cost of the analysis is reduced to a few CPU-minutes per source per detector.

The 5$n$-vector method uses a frequentist approach: the significance of a certain candidate, characterised by a value of the detection statistic, is established through the p-value, that is the probability to obtain a larger value for the statistics in the hypothesis of noise only. The statistic distribution can be inferred considering a range of off-source frequencies, as in \citet{narrband:2014}.
In case of no detection, upper limits on the amplitude for the single harmonic search are set following \citet{2019ApJ...879...10A}. For the dual harmonic search, for simplicity, we only consider the emission at once the rotation frequency, so we set upper limit on the $C_{21}$ parameter alone.

For the pulsars which experienced glitching, each inter-glitch period is analysed independently and then the resulting statistics are combined incoherently. 
For pulsars for which an estimation of the polarization parameters can be derived from EM observations, see Table \ref{tab:restricted_priors}, two searches are carried on: one assuming uninformative uniform priors on $\psi$ and $\iota$ and one using restricted Gaussian priors. 
Only O3 data from LIGO and Virgo detectors have been used in this search.

\section{Data sets used}
\label{sec:data}
\subsection{Gravitational-wave data}
\label{sec:data_gw}

The data set used O2 and O3 runs. The O2 run took place between 2016 October 30 (MJD: 57722.667) and 2017 August 25 (MJD: 57990.917). Virgo joined O2 on 2017 August 1. The duty factors for L1, H1, and V1, were $57\%$, $59\%$, and $80\%$, respectively. O3 took place between 2019 April 1 (MJD: 58574.625) and 2020 March 27 (MJD: 58935.708). Virgo was operational for the whole of the O3 run. The duty factors for this run were $76\%$, $71\%$, and $76\%$ for L1, H1, and V1, respectively. The uncertainties on the amplitude and phase calibration of the detectors for O2 can be found in \citet{2017PhRvD..96j2001C, 2018CQGra..35t5004A} and those for O3 can be found in \citet{2020CQGra..37v5008S, 2021arXiv210700129S, 2021arXiv210703294V}. For O2, the maximum $1\sigma$ amplitude uncertainties over the range 10-2000\,Hz were between about $[-2.5, +7.5]\%$ and $[-8, +4]\%$ for H1 and L1, respectively, and for Virgo the the maximum uncertainty was 5.1\%. For O3, the maximum $1\sigma$ amplitude uncertainties over the range 10-2000\,Hz were between about $[-5, +7]\%$ and $[-5.5, +5.5]\%$ for H1 and L1, respectively, and for Virgo the maximum uncertainty was 5\%. These ranges are the maximum upper and lower bound over the full frequency range and over different measurement epochs over the run, so at specific frequencies/times the uncertainty can be far smaller.

The data used underwent cleaning processes \citep{2019CQGra..36e5011D, cleaning, 2021arXiv210703294V}, specifically the removal of narrowband spectral artifacts at the calibration line frequencies and power line frequencies. A discussion on the consequences of performing a search using LIGO data with the narrowband cleaning of \citet{cleaning} applied compared to that without it applied can be found in Appendix \secref{appendix:app_comp}.

\subsection{Electromagnetic data}
\label{sec:data_em}

EM observations of pulsars produce the timing solutions used as input to the GW searches. These observations have been made in radio and X-ray wavelengths. The observatories which have contributed to the data set are: the Canadian Hydrogen Intensity Mapping Experiment (CHIME) \citep[as part of the CHIME Pulsar Project;][]{2021ApJS..255....5A}, the Mount Pleasant Observatory 26~m telescope, the 42~ft telescope and Lovell telescope at Jodrell Bank, the MeerKAT project \citep[as part of the MeerTime project;][]{MeerTime}, the Nan\c{c}ay Decimetric Radio Telescope, the Neutron Star Interior Composition Explorer (NICER) and the Molonglo Observatory Synthesis Telescope \citep[as part of the UTMOST pulsar timing programme;][]{2019MNRAS.484.3691J, 2020MNRAS.494..228L}.
Pulsar timing solutions were determined from this data using {\sc Tempo} \citep{2015ascl.soft09002N} or {\sc Tempo2} \citep{tempo22, tempo21, tempo23} to fit the model parameters.

We select pulsars whose rotation frequency is greater than 10 Hz so they are within the sensitivity band of the GW detectors. This leads to primarily targeting millisecond pulsars and fast spinning young pulsars. Of the 236 pulsars in this analysis, 74 are different from the 221 used in the O2 analysis \citep{2019ApJ...879...10A}. There are 168 pulsars in binary systems and 161 are millisecond pulsars with frequencies above about 100 Hz. The pulsar J0537\textminus6910 is not included due to the recently published searches for it in \citet{2021arXiv210414417T, 2021ApJ...913L..27A}.

For some pulsars, ephemerides were only available for the course of O3. In such cases, only GW data from O3 was used. This was the case for 102 out of the 236 pulsars in this analysis.

\section{Analysis results}
\label{sec:results}

No evidence for gravitational-wave signals from any of the included pulsars was found. The results for all except the high-value targets are shown in \tabref{tab:results}. The high-value pulsars are shown in \tabref{tab:highvalue} and discussed in Section~\secref{sec:high_value}. As no CWs were observed, we present the 95\% credible upper limits on the gravitational-wave amplitudes $C_{22}$ and $C_{21}$ for the dual harmonic run (searching for the mass quadrupole modes $l=2$, $m=1,2$) and the gravitational-wave amplitude $h_0$ for the single harmonic ($l=2$, $m=2$) search. These were all calculated using coherently combined data from all three detectors over the O2 and O3 observing runs or just the O3 run, as appropriate. Due to the calibration amplitude systematic uncertainties for the detectors, the amplitude estimates can have uncertainties of up to $\sim8\%$.

\begin{figure*}
\epsscale{1.05}
\plotone{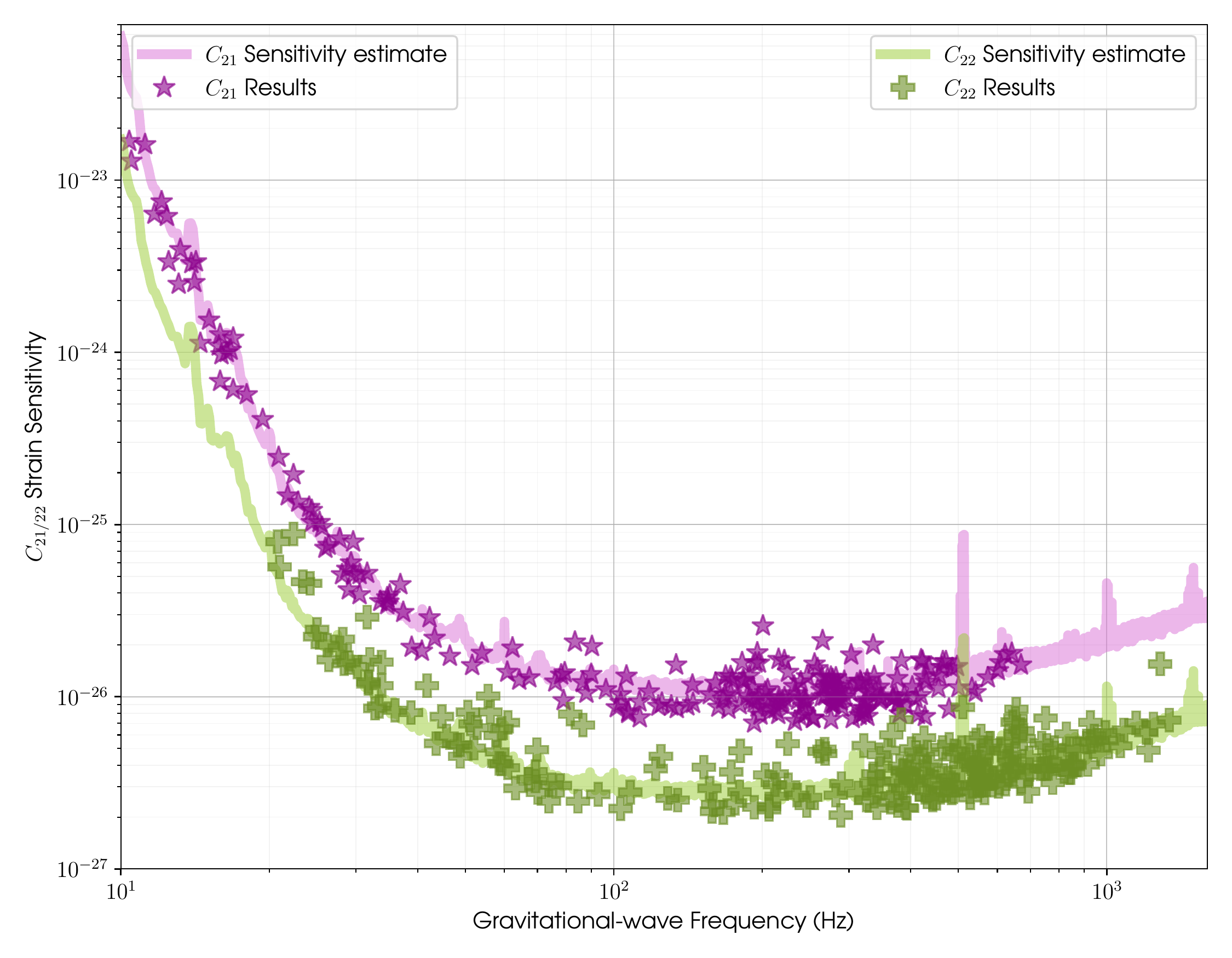}
\caption{The 95\% credible upper limits on the gravitational-wave amplitudes for all 236 pulsars using the time-domain Bayesian method. The pink stars and green crosses show the 95\% credible upper limits for the GW amplitudes ($C_{22}$ and $C_{21}$) for the dual harmonic search. The solid lines show the estimated sensitivity of all three detectors combined over the course of O3.}
\label{fig:O3Csenscurve}
\end{figure*}

\figref{fig:O3Csenscurve} shows the 95\% credible upper limits on the gravitational-wave amplitudes $C_{22}$ and $C_{21}$ for all pulsars using the time-domain Bayesian method. In addition, it shows joint detector sensitivity estimates for the two amplitudes based on the representative power spectral densities for the detectors over the course of O3. For an explanation of how these estimates were calculated, see Appendix~C in \citet{2019ApJ...879...10A}.

\begin{figure*}
\epsscale{1.05}
\plotone{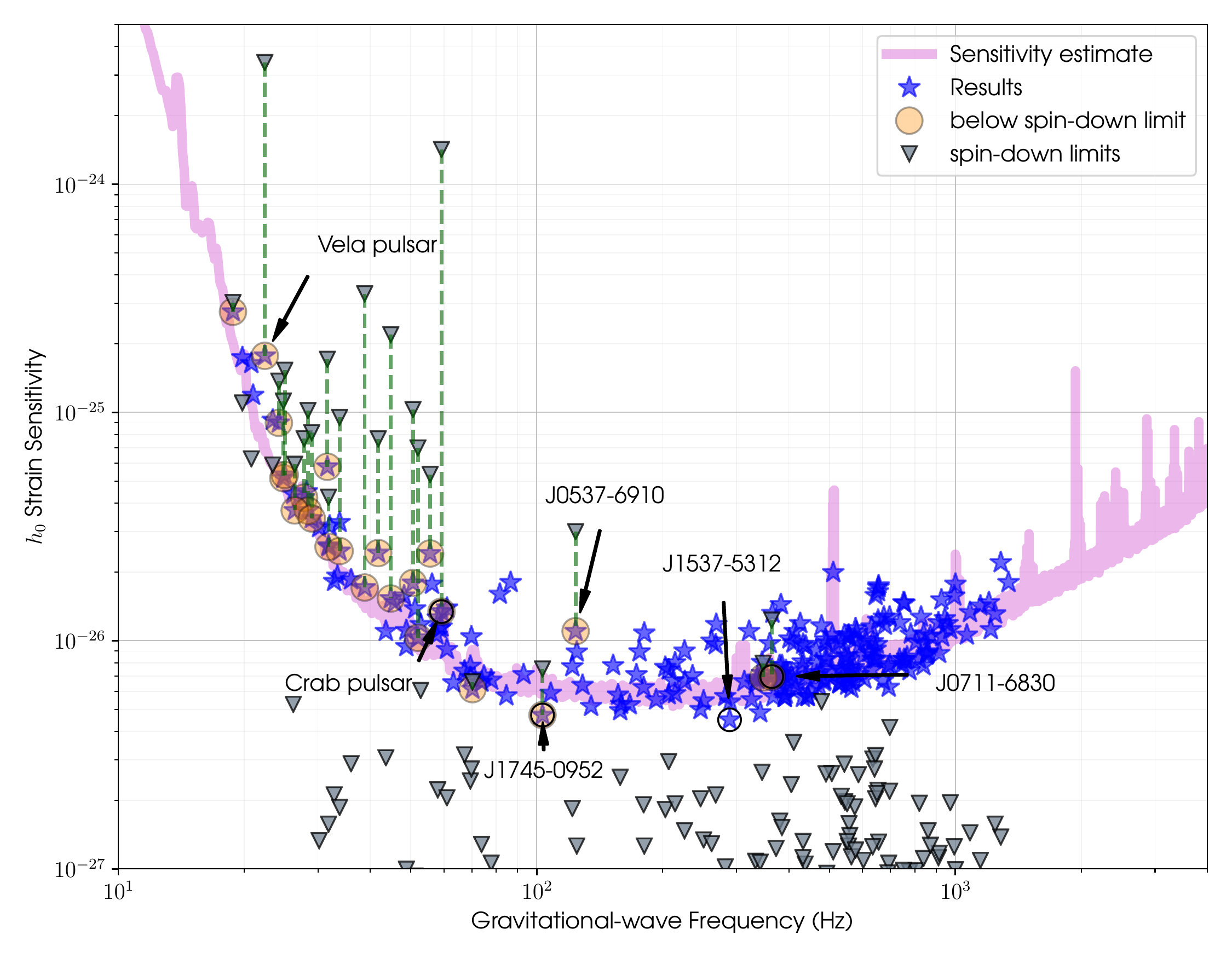}
\caption{Upper limits on $h_0$ for the 236 pulsars in this analysis using the time-domain Bayesian method. The stars show 95\% credible upper limits on the amplitudes of $h_0$. Grey triangles represent the spin-down limits for each pulsar (based on the distance measurement stated in \tabref{tab:results} and assuming the canonical moment of inertia). For those pulsars which surpass their spin-down limits, their results are plotted within shaded circles. The pink curve gives an estimate of the expected strain sensitivity of all three detectors combined during the course of O3. The highlighted pulsars are those with the best $h_0$, $Q_{22}$ and spin-down ratio out of the pulsars which surpassed their spin-down limit, as well as the best $h_{0}$ limit out of the whole sample. The Vela pulsar is highlighted and the pulsar J0537\textminus6910 upper and spin-down limits calculated in \cite{2021ApJ...913L..27A} are also included for completeness.}
\label{fig:h0senscurve}
\end{figure*}

The 95\% credible upper limits for the GW amplitude $h_0$ from the single harmonic analysis for all pulsars, again using the results from the time-domain Bayesian method, are shown as stars in \figref{fig:h0senscurve}. The spin-down limit for each pulsar is represented as a grey triangle. If the observed upper limit for a pulsar is below the spin-down limit, this is shown via a dotted green line from the spin-down limit to the $h_0$ limit which is plotted within a shaded circle. The solid line gives the joint detector sensitivity estimate over the course of O3.\footnote{The sensitivity estimate for O3 alone is used as it dominates compared to the estimate for the O2 run.}

\begin{figure}
\epsscale{1.15}
\plotone{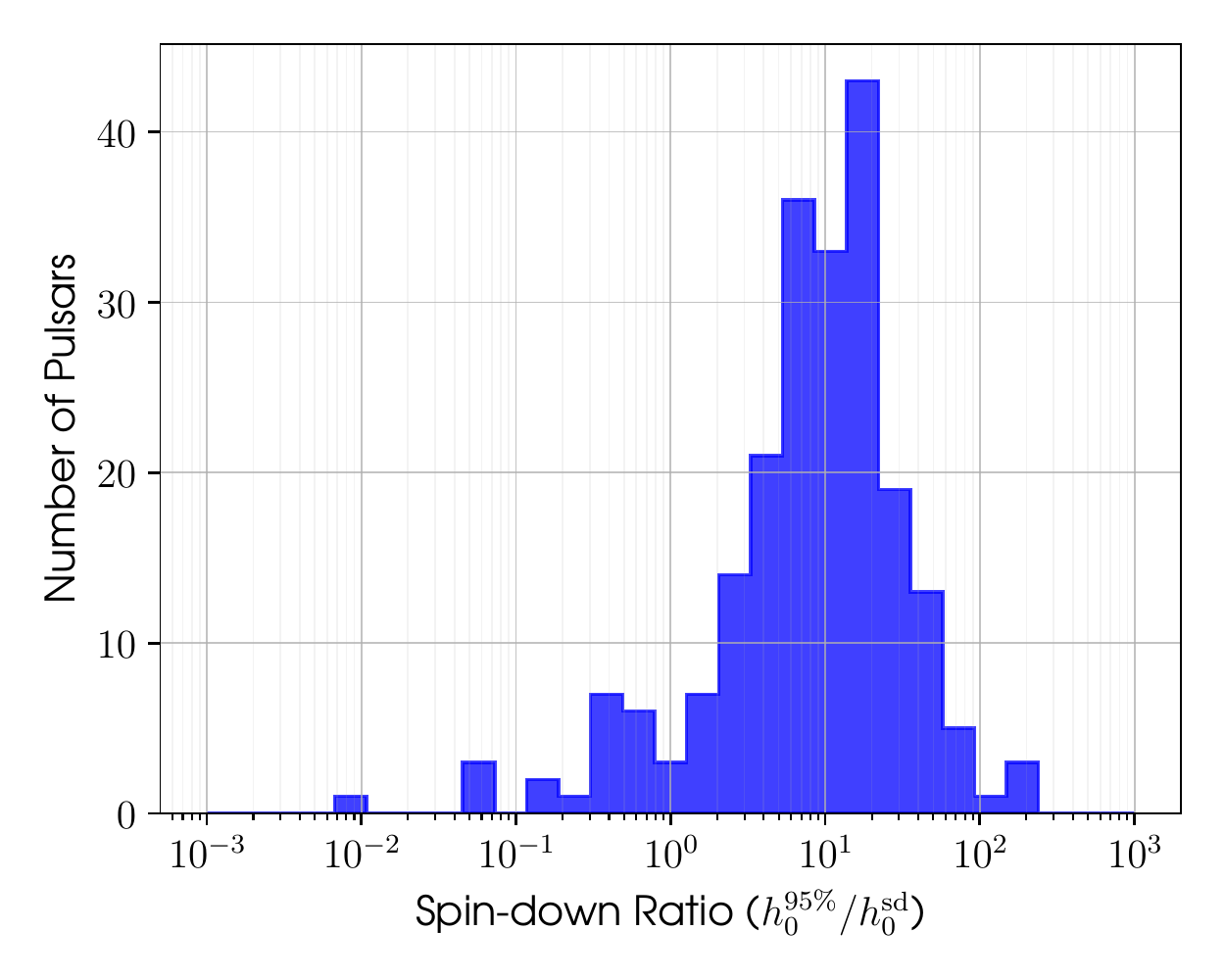}
\caption{A histogram of the spin-down ratio for all pulsars for which a spin-down ratio was calculated.}
\label{fig:ratiohist}
\end{figure}

\figref{fig:ratiohist} shows a histogram of the spin-down ratio $h_0^{95\%}/h_0^{\rm sd}$ from the Bayesian analysis for every pulsar for which calculating a spin-down rate was possible.\footnote{Spin-down rates cannot be calculated for pulsars with insufficient distance, frequency or frequency derivative data (see equation~\eqref{eq:spin-down}).} These values rely on the pulsar distance, frequency derivative and principal moment of inertia, which all have associated uncertainties. These are not taken into account in this study, for which we use the best-fit values listed in \tabref{tab:results} and \ref{tab:highvalue} and a fiducial moment of inertia $I^{\rm fid}_{zz}$ of $10^{38}\,{\rm kg}\,{\rm m}^2$, but their presence should be kept in mind. Distance errors are primarily based on uncertainties in the galactic free electron distribution \citep{2017ApJ...835...29Y}, which can lead to distance errors on the order of a factor of two. Nearby pulsars, for which parallax measurements are possible, will generally have smaller distance uncertainties. \tabref{tab:results} provides a reference for the distance to each pulsar, which can be used to find an estimate of the associated error as required. The relative uncertainties in frequency derivative are generally much smaller than the distance uncertainties. The principal moment of inertia is equation of state dependent and could range between $\sim (1-3)\!\times\!10^{38}\,{\rm kg}\,{\rm m}^2$ \citep[see, e.g.,][]{2007PhRvD..76d2001A}. The mass quadrupole $Q_{22}$ and the ellipticity $\varepsilon$ limits also rely on these values; for example, our given ellipticity upper limits are inversely proportional to $I_{zz}/10^{38}\,{\rm kg}\,{\rm m}^2$. 

\begin{figure*}
\epsscale{1.05}
\plotone{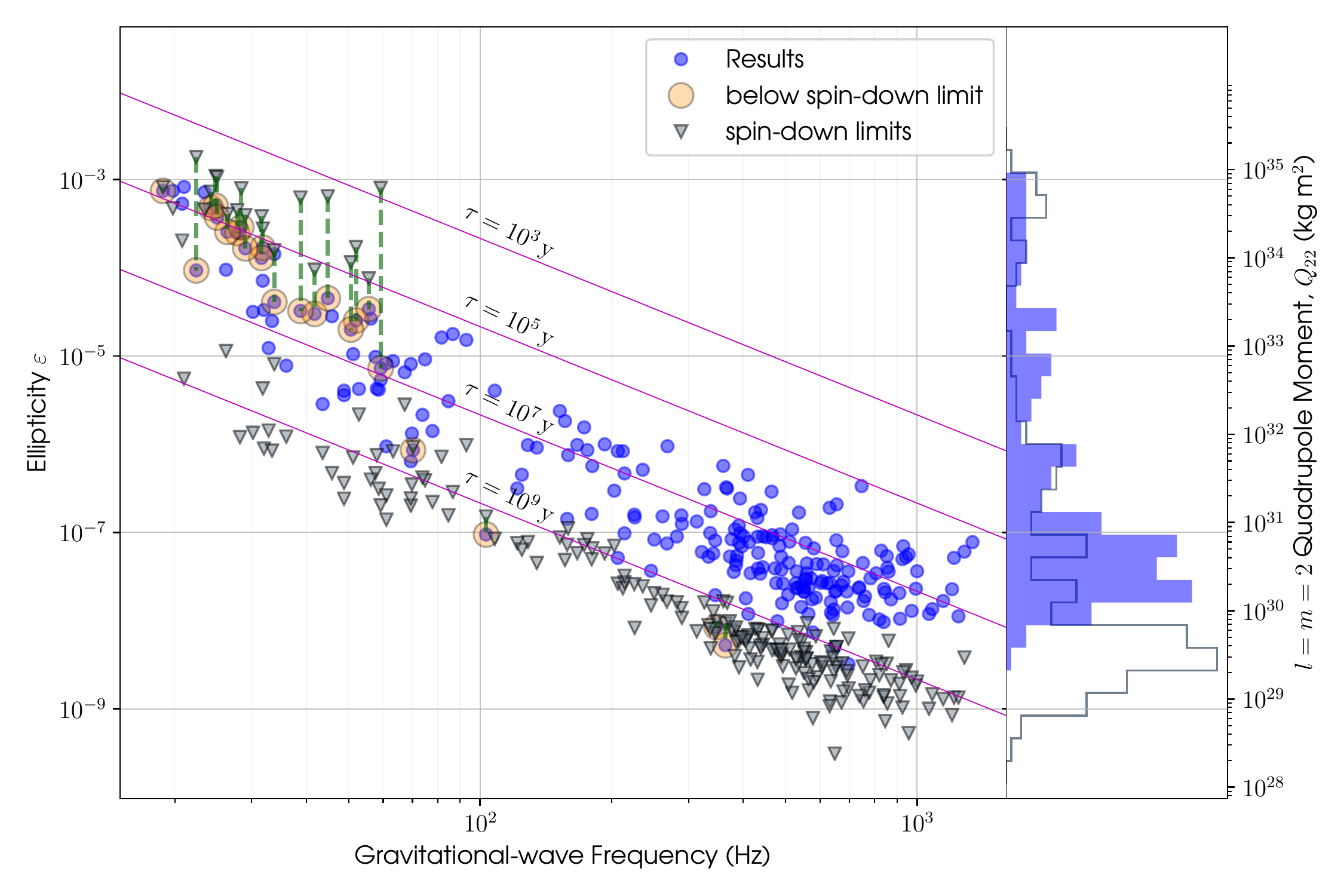}
\caption{95\% credible upper limits on ellipticity $\varepsilon$ and mass quadrupole $Q_{22}$ for all 236 pulsars. The upper limits for each pulsar are represented by blue circles while their spin-down limits are shown as grey triangles. Pulsars for which our direct upper limits have surpassed their spin-down limits are highlighted within a shaded circle with a dotted green line linking the limit to its spin-down limit. Also included are pink contour lines of equal characteristic age $\tau = P/4\dot{P}$ assuming that GW emission alone is causing spin-down. To the right of the plot, histograms of both these direct limits and spin-down limits are shown by filled and empty bars respectively.}
\label{fig:ellfreq}
\end{figure*}

The single harmonic search was used to place limits on the mass quadrupole $Q^{95\%}_{22}$
which can be used to find the pulsar's ellipticity $\varepsilon^{95\%}$ using equations~\eqref{eq:h0} and \eqref{eq:q22}. However, for pulsars that did not surpass their spin-down limits these $Q_{22}$ and $\varepsilon$ values would lead to spin-down rates $\dot{P}_{rot}$ that are greater than (and thus are in violation of) their measured values.
The results for the Bayesian analysis are shown in terms of the mass quadrupole $Q_{22}$ and ellipticity $\varepsilon$ in \figref{fig:ellfreq}. Also included are histograms of the upper limits and spin-down limits as well as contour lines of equal characteristic age $\tau$ calculated under the assumption that all spin-down is due to energy loss through GW emission, i.e., the braking index is $n=5$.

From the Bayesian search twenty-three pulsars have direct upper limits that are below their spin-down limit, with 89 pulsars within a factor of 10 of their spin-down limit. There were 90 millisecond pulsars with a spin-down ratio less than 10. For the dual harmonic search, the most constraining upper limit for $C_{21}$ was J2302+4442 with $\scinum{7.05}{-27}$. The smallest $C_{22}$ upper limit was \scinum{2.05}{-27} for J1537\textminus5312. As physically meaningful constraints for the single harmonic search are only achieved once the spin-down limit has been surpassed, the following best limits are taken from the 23 pulsars that had $h_0^{95\%}/h_0^{sd} < 1$. The smallest spin-down limit was 0.009 for J0534+2200 (the Crab pulsar). The pulsar with the smallest upper limit on $h_0$ was J1745\textminus0952 with $\scinum{4.72}{-27}$. The best $Q_{22}$ upper limit was achieved by J0711\textminus6830 with \scinum{4.07}{29}\,kg\,m$^2$ which led to the best limit on ellipticity of \scinum{5.26}{-9}. This pulsar has a dispersion measure distance of 0.11\,kpc, which makes it relatively close by. However, its high ecliptic latitude makes it very insensitive to parallax measurement \citep{2021MNRAS.507.2137R}.

For each pulsar, we performed a model comparison between the assumption of the data being consistent with a coherent signal compared to the assumption of an incoherent signal {\it or} noise. This was calculated for both the dual harmonic ($l=2$, $m=1,2$) and single harmonic ($l=2$, $m=2$) searches. Specifically, the base-10 logarithm of the Bayesian odds between models is calculated, which will be referred to as $\mathcal{O}$.
Of all the pulsars in this search, none had $\mathcal{O} > 0$, meaning in all cases incoherent noise was more likely than a coherent signal. The pulsar with the highest odds overall was J2010\textminus1323 with $-0.77$.

\subsection{High-value targets}
\label{sec:high_value}

\tabref{tab:highvalue} shows the results for the analyses on the high-value targets for the \textit{Bayesian}, the \textit{$\mathcal{F}$/$\mathcal{G}$-statistic} and the \textit{5$n$-vectors} analyses. In this case, the two columns named ``Statistic" have different values depending on which analysis method was used. For the Bayesian analysis they give the base-10 logarithm of the Bayesian odds, $\mathcal{O}$, for the dual- and single-harmonic searches, respectively. This is the same as for the results in \tabref{tab:results}. 
For the \textit{$\mathcal{F}$/$\mathcal{G}$-statistic} and \textit{5$n$-vector} analysis methods they represent the false-alarm probabilities at the $l=2, m=1$ and $l=m=2$ modes respectively.

By definition, all high-value pulsars surpassed their spin-down limits in the Bayesian analysis. Several pulsars glitched during the course of the runs: J0534+2200 (Crab pulsar), J0908\textminus4913 and J1105\textminus6107. The times of the glitches are shown in \tabref{tab:glitches} and the process for dealing with them is outlined in Section~\secref{sec:glitches}. Additionally, some have sufficient information from EM observations on their orientation to restrict their priors: J0534+2200 (Crab pulsar), J0835\textminus4510 (Vela), J1952+3252 and J2229+6114. This is discussed in \secref{sec:restricted} and the pulsars' restricted ranges are quoted in \tabref{tab:restricted_priors}.
For each pulsar with either a glitch or restricted priors, individual analyses were performed assuming GW emission at both $2f_{\rm rot}$ and $f_{\rm rot}$ and just $2f_{\rm rot}$. In the case of the Crab pulsar, which both experienced a glitch and has sufficient observations for restricted priors, four individual analyses were performed. Each analysis accounted for the glitch, with combinations of dual/single harmonic search and restricted/unrestricted priors. The values shown in \tabref{tab:highvalue} are from the searches with glitches accounted for via an unknown phase offset. When a pulsar had a restricted prior search, the results are shown in parentheses next to the unrestricted results. 

The Crab pulsar is of interest due to its high spin-down luminosity. For the single harmonic Bayesian analysis and with the glitch accounted for by a phase offset, its upper limit as a fraction of the spin-down limit is only $0.0094(0.0085)$ meaning that GWs contribute to less than 0.009\% of the available spin-down luminosity. This is consistent with previous studies that also surpassed the spin-down limit \citep{2019ApJ...879...10A,2017PhRvD..96l2006A}. Its $h_0^{95\%}$ upper limit was found to be \scinum{1.3(1.2)}{-26}. With a distance of 2\,kpc and period derivative of \scinum{4.2}{-13}\,s\,s$^{-1}$, the upper limits on the mass quadrupole and ellipticity were calculated to be $Q_{22}^{95\%}=\scinum{5.6(5.0)}{32}$ kg m$^2$ and $\varepsilon^{95\%}=\scinum{7.2(6.5)}{-6}$. The base-10 logarithm of the Bayesian odds for this analysis favouring a coherent signal over incoherent noise is -2.6(-2.7). 

The Vela pulsar also has a very high spin-down luminosity and is considered another source of interest. Unlike the Crab pulsar, the Vela pulsar did not experience any glitches over the course of this analysis. In the single harmonic Bayesian analysis, the spin-down limit was surpassed with a ratio of 0.052(0.051), with $h_0^{95\%} = \scinum{1.8(1.7)}{-25}$. This ratio corresponds to a maximum of 0.27\% of the spin-down luminosity being emitted by GWs. The previous known pulsar search by \citet{2019ApJ...879...10A} found the spin-down ratio to be 0.042 with $h_0^{95\%} = \scinum{1.4}{-25}$ which is lower than in this analysis. This is due to significant noise in the LIGO Hanford detector at twice Vela's rotational frequency, with an angular sensing control dither line being the most likely culprit.\footnote{This contamination was removed for the final third of O3, although its presence at earlier times still has a detrimental effect on the result.} However, this analysis is an improvement on the more recent measurement of the spin-down ratio of 0.067 and $h_0^{95\%} = \scinum{2.2}{-25}$ produced in \citet{Abbott_2020}.
The upper limits on the mass quadrupole $Q_{22}^{95\%}$ and ellipticity $\varepsilon$ were calculated to be \scinum{7.2(7.1)}{33} kg m$^2$ and \scinum{9.3(9.2)}{-5}, respectively. These values were calculated assuming a distance of 0.28\,kpc and a period derivative of \scinum{1.2}{-13}\,s\,s$^{-1}$. The base-10 logarithm of the Bayesian odds for this pulsar in the single harmonic analysis was -1.1(-1.0). 

The pulsar J0537\textminus6910 has the highest spin-down luminosity but has not been included in this search due to recently published searches for it in \citet{2021ApJ...913L..27A, 2021arXiv210414417T}. The limits, which can be found in Table 3 of \citet{2021ApJ...913L..27A}, are shown for comparison. They found $h_0^{95\%}=\scinum{1.1(1.0)}{-26}$ with a spin-down ratio of $h_0^{95\%}/h_0^{sd}=0.37(0.33)$. $\varepsilon^{95\%}=\scinum{3.4(3.1)}{-5}$ while for the dual harmonic search $C_{21}^{95\%}=\scinum{2.2(1.8)}{-26}$ and $C_{22}^{95\%}=\scinum{5.6(5.0)}{-27}$. 

Table~\ref{tab:BDresults} shows the results for the analyses on the high-value targets using the $\mathcal{D}$-statistic analyses to search for dipole radiation predicted by Brans-Dicke theory. No significant signal was detected and consequently upper limits are obtained. The 95\% confidence upper limits are given in the second last column of Table~\ref{tab:BDresults} and the last column gives the false alarm probability,  i.e., the probability that the obtained values of the $\mathcal{D}$-statistic result only from the noise in the data. The most constraining upper limit for dipole radiation is obtained for the millisecond pulsar J0437\textminus4715.

\section{Conclusions}
\label{sec:conclusions}

In this paper, we have searched for evidence of GWs from 236 pulsars over the course of the LIGO and Virgo O2 and O3 runs and across all three detectors (LIGO Hanford, LIGO Livingston and Virgo). This is an improvement on the 221 pulsars from the O1 and O2 analysis in \citet{2019ApJ...879...10A}. Searches were carried out for two different emission models. One assumed GW emission from the $l=m=2$ mass quadrupole mode and the other assumed emission from the $l=2, m=1,2$ modes. For the single harmonic search, new upper limits on $h_0$ were produced and a total of 23 pulsars surpassed their spin-down limits (24 if one includes J0537-6910 from \citet{2021ApJ...913L..27A}). This is an improvement from the 20 pulsars in \citet{2019ApJ...879...10A} and includes 9 pulsars for which their spin-down limit had not previously been surpassed. For the dual harmonic search, new limits on $C_{21}$ and $C_{22}$ are found. For pulsars which were deemed high-value, two more analysis methods were included for robustness: the $\mathcal{F}$/$\mathcal{G}$-\textit{statistic} method and the \textit{5n-vector} method. 

The millisecond pulsars that surpassed their spin-down limits, J0437\textminus4715 and J0711\textminus6830, have ellipticity upper limits of $\scinum{8.5}{-9}$ and $\scinum{5.3}{-9}$, respectively. Comparing these values to the left-hand panel of Figure~3 in \citet{gittins2021modelling} finds that our results are lower than the maximum values predicted for a variety of neutron star equations of state. Our results are therefore providing new constraints in physically realistic parts of the ellipticity parameter space.

No search found strong evidence of GW emission from any of the pulsars. However, with so many pulsars now surpassing their spin-down limit, including the millisecond pulsars J0437\textminus4715 and J0711\textminus6830 \citep{Abbott_2020}, the next observing run O4 could add more pulsars to this count and bring us closer to observing CWs from pulsars for the first time. In addition to the search for CW signals consisting of the tensorial polarizations predicted by GR, this paper provides the first search explicitly targeting emission of scalar polarization modes predicted by Brans-Dicke theory.

\section{Acknowledgements}
\label{sec:acknowledgements}
This material is based upon work supported by NSF’s LIGO Laboratory which is a major facility
fully funded by the National Science Foundation.
The authors also gratefully acknowledge the support of
the Science and Technology Facilities Council (STFC) of the
United Kingdom, the Max-Planck-Society (MPS), and the State of
Niedersachsen/Germany for support of the construction of Advanced LIGO 
and construction and operation of the GEO600 detector. 
Additional support for Advanced LIGO was provided by the Australian Research Council.
The authors gratefully acknowledge the Italian Istituto Nazionale di Fisica Nucleare (INFN),  
the French Centre National de la Recherche Scientifique (CNRS) and
the Netherlands Organization for Scientific Research (NWO), 
for the construction and operation of the Virgo detector
and the creation and support  of the EGO consortium. 
The authors also gratefully acknowledge research support from these agencies as well as by 
the Council of Scientific and Industrial Research of India, 
the Department of Science and Technology, India,
the Science \& Engineering Research Board (SERB), India,
the Ministry of Human Resource Development, India,
the Spanish Agencia Estatal de Investigaci\'on (AEI),
the Spanish Ministerio de Ciencia e Innovaci\'on and Ministerio de Universidades,
the Conselleria de Fons Europeus, Universitat i Cultura and the Direcci\'o General de Pol\'{\i}tica Universitaria i Recerca del Govern de les Illes Balears,
the Conselleria d'Innovaci\'o, Universitats, Ci\`encia i Societat Digital de la Generalitat Valenciana and
the CERCA Programme Generalitat de Catalunya, Spain,
the National Science Centre of Poland and the European Union – European Regional Development Fund; Foundation for Polish Science (FNP),
the Swiss National Science Foundation (SNSF),
the Russian Foundation for Basic Research, 
the Russian Science Foundation,
the European Commission,
the European Social Funds (ESF),
the European Regional Development Funds (ERDF),
the Royal Society, 
the Scottish Funding Council, 
the Scottish Universities Physics Alliance, 
the Hungarian Scientific Research Fund (OTKA),
the French Lyon Institute of Origins (LIO),
the Belgian Fonds de la Recherche Scientifique (FRS-FNRS), 
Actions de Recherche Concertées (ARC) and
Fonds Wetenschappelijk Onderzoek – Vlaanderen (FWO), Belgium,
the Paris \^{I}le-de-France Region, 
the National Research, Development and Innovation Office Hungary (NKFIH), 
the National Research Foundation of Korea,
the Natural Science and Engineering Research Council Canada,
Canadian Foundation for Innovation (CFI),
the Brazilian Ministry of Science, Technology, and Innovations,
the International Center for Theoretical Physics South American Institute for Fundamental Research (ICTP-SAIFR), 
the Research Grants Council of Hong Kong,
the National Natural Science Foundation of China (NSFC),
the Leverhulme Trust, 
the Research Corporation, 
the Ministry of Science and Technology (MOST), Taiwan,
the United States Department of Energy,
and
the Kavli Foundation.
The authors gratefully acknowledge the support of the NSF, STFC, INFN and CNRS for provision of computational resources.

This work was supported by MEXT, JSPS Leading-edge Research Infrastructure Program, JSPS Grant-in-Aid for Specially Promoted Research 26000005, JSPS Grant-in-Aid for Scientific Research on Innovative Areas 2905: JP17H06358, JP17H06361 and JP17H06364, JSPS Core-to-Core Program A. Advanced Research Networks, JSPS Grant-in-Aid for Scientific Research (S) 17H06133 and 20H05639 , JSPS Grant-in-Aid for Transformative Research Areas (A) 20A203: JP20H05854, the joint research program of the Institute for Cosmic Ray Research, University of Tokyo, National Research Foundation (NRF) and Computing Infrastructure Project of KISTI-GSDC in Korea, Academia Sinica (AS), AS Grid Center (ASGC) and the Ministry of Science and Technology (MoST) in Taiwan under grants including AS-CDA-105-M06, Advanced Technology Center (ATC) of NAOJ, Mechanical Engineering Center of KEK.

The Nan\c{c}ay Radio Observatory is operated by the Paris Observatory, associated with the French Centre National de la Recherche Scientifique (CNRS). We acknowledge financial support from ``Programme National de Cosmologie et Galaxies’’ (PNCG) and ``Programme National Hautes Energies’’ (PNHE) of CNRS/INSU, France. 

We are grateful to the staff of the Dominion Radio Astrophysical Observatory, which is operated by the National Research Council of Canada.  CHIME is funded by a grant from the Canada Foundation for Innovation (CFI) 2012 Leading Edge Fund (Project 31170) and by contributions from the provinces of British Columbia, Qu\'{e}bec and Ontario. The CHIME/FRB Project, which enabled development in common with the CHIME/Pulsar instrument, is funded by a grant from the CFI 2015 Innovation Fund (Project 33213) and by contributions from the provinces of British Columbia and Qu\'{e}bec, and by the Dunlap Institute for Astronomy and Astrophysics at the University of Toronto. Additional support was provided by the Canadian Institute for Advanced Research (CIFAR), McGill University and the McGill Space Institute thanks to the Trottier Family Foundation, and the University of British Columbia. The CHIME/Pulsar instrument hardware was funded by NSERC RTI-1 grant EQPEQ 458893-2014. This research was enabled in part by support provided by WestGrid (www.westgrid.ca) and Compute Canada (www.computecanada.ca). J.W.M.\ is a CITA Postdoctoral Fellow: This work was supported by the Natural Sciences and Engineering Research Council of Canada (NSERC), funding reference \#CITA 490888-16. Pulsar research at UBC is supported by an NSERC Discovery Grant and by the Canadian Institute for Advanced Research (I.H.S.). K.C.\ is supported by a UBC Four Year Fellowship (6456).

D.An.\ acknowledges support from an EPSRC/STFC fellowship (EP/T017325/1). W.C.G.H.\ acknowledges support through grants 80NSSC19K1444 and 80NSSC21K0091 from NASA. This work is supported by NASA through the NICER mission and the Astrophysics Explorers Program and uses data and software provided by the High Energy Astrophysics Science Archive Research Center (HEASARC), which is a service of the Astrophysics Science Division at NASA/GSFC and High Energy Astrophysics Division of the Smithsonian Astrophysical Observatory.

We would like to thank all of the essential workers who put their health at risk during the COVID-19 pandemic, without whom we would not have been able to complete this work.

\software{
\code{Astropy}~\citep{astropy:2013,astropy:2018},
\code{GWpy}~\citep{gwpy},
\code{LALSuite}~\citep{LALSuite},
\code{matplotlib}~\citep{2007CSE.....9...90H},
\code{numpy}~\citep{harris2020array},
\code{psrqpy}~\citep{psrqpy},
\code{scipy}~\citep{scipy},
\code{SWIGLAL}~\citep{Wette:2020air},
\code{Tempo}~\citep{2015ascl.soft09002N},
\code{Tempo2}~\citep{2006MNRAS.369..655H}.
}

\appendix

\section{Cleaned versus uncleaned data comparison}
\label{appendix:app_comp}

The data used in this analysis was subject to a cleaning process described in \citet{cleaning} which focused on the removal of various narrow-band spectral artifacts at calibration line frequencies. For any pulsars very close to these lines, this cleaning would be expected to provide an improvement in sensitivity. In this appendix we present the comparison of results using this cleaned data against results using data without this cleaning process (which we will refer to as ``uncleaned'') for a sample of pulsars.

\begin{figure*}
\epsscale{1.05}
\plotone{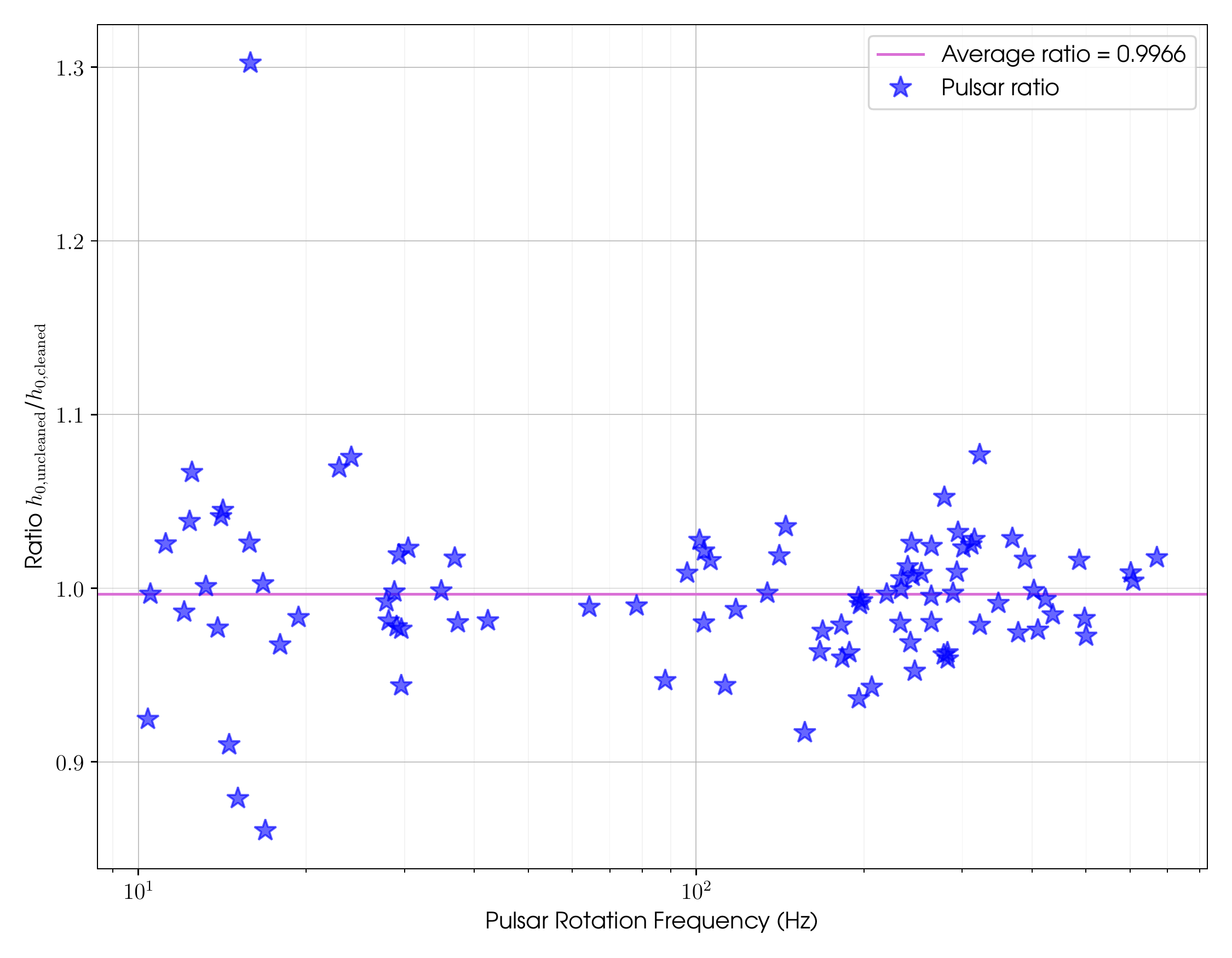}
\caption{The ratio of uncleaned versus cleaned upper limits against pulsar rotational frequency. The $h_0$ upper limits for both data sets are calculated from the $C_{22}$ limit determined in dual harmonic searches using O3 data only. The mean ratio is represented by the solid pink line.}
\label{fig:comparison}
\end{figure*}

Uncleaned O3 LIGO data is used for a dual harmonic Bayesian analysis of 95 pulsars which had ephemeris data only overlapping with O3. This is compared to the O3-only analysis performed in this paper using the cleaned data. The Virgo data used was the same in both cases. For comparison, the ratio of $h_0$ upper limits for each pulsar using uncleaned $h_{0,\textrm{uncleaned}}$ versus cleaned $h_{0,\textrm{cleaned}}$ data are shown in \figref{fig:comparison}.

The mean ratio of upper limit for uncleaned data versus cleaned data was 0.9966 (with a standard deviation of 0.0486) which suggests no major effect of the line cleaning on the majority results. It should be noted that for this analysis there is a statistical uncertainty on the upper limits of around 1\% due to the use of a finite number of posterior samples when calculating them \citep{Abbott_2020}. If performing parameter estimation on $h_0$ using multiple data sets consisting of independent noise realizations drawn from the same distribution, empirically it is found that the resultant upper limits will vary by on order of 30\%. Due to the cleaning, the cleaned and uncleaned datasets will contain different, albeit highly correlated, noise. So, a spread of upper limit ratios that are larger than expected from the pure statistical uncertainty on each limit, but smaller that one would get from independent data is to be expected.

As the upper limit ratio spread can be explained as being consistent with expectations from statistical fluctuations, it suggests that very few pulsars are close enough to the cleaned lines for the cleaning to have a significant effect overall. However, it makes sense to apply consistency in using the same dataset for all pulsars being analyzed. In this analysis we chose to use the narrow-band cleaned data.

\bibliographystyle{aasjournal}
\bibliography{bibliography}

\startlongtable
\begin{longrotatetable}
\begin{deluxetable}{lrlllhllllllll}
\tablecaption{Limits on Gravitational-wave Amplitude, and Other Derived Quantities, for 213  Pulsars from the {\it Bayesian} Analysis.\label{tab:results}}
\tabletypesize{\footnotesize}
\tablehead{
\colhead{Pulsar Name} & 
\colhead{$f_{\text{rot}}$} & 
\colhead{$\dot{P}_{\text{rot}}$} & 
\colhead{Distance} & 
\colhead{$h_0^{\text{sd}}$} & 
\nocolhead{Analysis} & 
\colhead{$C_{21}^{95\%}$} & 
\colhead{$C_{22}^{95\%}$} & 
\colhead{$h_0^{95\%}$} & 
\colhead{$Q_{22}^{95\%}$} & 
\colhead{$\varepsilon^{95\%}$} & 
\colhead{$h_0^{95\%}/h_0^{\text{sd}}$} & 
\colhead{$\mathcal{O}^{l=2}_{m=1,2}$} & 
\colhead{$\mathcal{O}^{l=2}_{m=2}$}\\ 
\colhead{(J2000)} & 
\colhead{(Hz)} & 
\colhead{(s\,s$^{-1})$} & 
\colhead{(kpc)} & 
\colhead{~} & 
\nocolhead{Method} & 
\colhead{~} & 
\colhead{~} & 
\colhead{~} & 
\colhead{(kg\,m$^2$)} & 
\colhead{~} & 
\colhead{~} & 
\colhead{~} & 
\colhead{~}}
\startdata
J0023+0923\tablenotemark{a}\tablenotemark{$\epsilon$} & 327.8 & \scinum{1.0}{-20} & 1.11 (a) & \scinum{1.3}{-27} & Bayesian & \scinum{9.6}{-27} & \scinum{8.5}{-27} & \scinum{1.7}{-26} & \scinum{3.3}{30} & \scinum{4.2}{-8} & 13 & -3.2 & -1.0 \\ 
J0030+0451\tablenotemark{a}\tablenotemark{$\epsilon$} & 205.5 & \scinum{1.0}{-20} & 0.33 (a) & \scinum{3.6}{-27} & Bayesian & \scinum{8.7}{-27} & \scinum{3.1}{-27} & \scinum{6.5}{-27} & \scinum{9.2}{29} & \scinum{1.2}{-8} & 1.8 & -4.4 & -2.3 \\ 
J0034\textminus0534\tablenotemark{a}\tablenotemark{$\epsilon$} & 532.7 & \scinum{4.2}{-21} & 1.35 (b) & \scinum{8.9}{-28} & Bayesian & \scinum{1.1}{-26} & \scinum{5.2}{-27} & \scinum{1.1}{-26} & \scinum{9.2}{29} & \scinum{1.2}{-8} & 12 & -4.7 & -2.4 \\ 
J0101\textminus6422\tablenotemark{a}\tablenotemark{$\delta$} & 388.6 & \scinum{3.7}{-21} & 1.00 (b) & \scinum{9.7}{-28} & Bayesian & \scinum{1.0}{-26} & \scinum{3.2}{-27} & \scinum{7.4}{-27} & \scinum{9.0}{29} & \scinum{1.2}{-8} & 7.6 & -4.7 & -2.5 \\ 
J0102+4839\tablenotemark{$\epsilon$} & 337.4 & \scinum{1.1}{-20} & 2.31 (b) & \scinum{6.8}{-28} & Bayesian & \scinum{1.3}{-26} & \scinum{6.3}{-27} & \scinum{1.3}{-26} & \scinum{4.7}{30} & \scinum{6.1}{-8} & 19 & -3.7 & -1.6 \\ 
J0117+5914\tablenotemark{a}\tablenotemark{$\alpha$} & 9.9 & \scinum{5.9}{-15} & 1.77 (b) & \scinum{1.1}{-25} & Bayesian & \nodata & \nodata & \scinum{1.7}{-25} & \scinum{5.8}{34} & \scinum{7.5}{-4} & 1.6 & \nodata & -2.5 \\ 
J0125\textminus2327\tablenotemark{$\epsilon$} & 272.1 & \scinum{1.8}{-20} & 0.92 (b) & \scinum{2.0}{-27} & Bayesian & \scinum{1.2}{-26} & \scinum{4.2}{-27} & \scinum{8.5}{-27} & \scinum{1.9}{30} & \scinum{2.5}{-8} & 4.3 & -4.1 & -2.1 \\ 
J0154+1833\tablenotemark{a}\tablenotemark{$\epsilon$} & 422.9 & \scinum{1.1}{-21} & 1.62 (b) & \scinum{3.4}{-28} & Bayesian & \scinum{1.6}{-26} & \scinum{3.6}{-27} & \scinum{9.7}{-27} & \scinum{1.6}{30} & \scinum{2.1}{-8} & 29 & -4.3 & -2.5 \\
\enddata
\tablerefs{The following is a list of references for pulsar distances and intrinsic period derivatives, and they should be consulted for information on the associated uncertainties on these quantities: (a) \citet{2018ApJS..235...37A}, (b) \citet{2017ApJ...835...29Y}, (c) \citet{2014MNRAS.444.1859V}, (d) \citet{2013Sci...340..448A}, (e) \citet{2016MNRAS.455.1751R}, (f) \citet{2021MNRAS.507.2137R}, (g) \citet{2016MNRAS.455.3806B}, (h) \citet{2021ApJ...915L..12F}, (i) \citet{2016MNRAS.458.3341D}, (j) \citet{2012ApJ...755...39V}, (k) \citet{2021arXiv210803946L}, (l) \citet{2006MNRAS.369..416R}, (m) \citet{2011ApJ...732L..11N}, (n) \citet{2017ApJ...847...25S}, (o) \citet{2013ApJ...778..120H}, (p) \citet{2015MNRAS.446.4019B}, (q) \citet{2016ApJ...818...92M}, (r) \citet{2010arXiv1012.3224H}, (s) \citet{2014ApJ...787...82F}, (t) \citet{2015ApJ...799..165B}, (u) \citet{2018ApJ...855..122V}, (v) \citet{2012MNRAS.423.3328F}, (w) \citet{2013MNRAS.430..571E}, (x) \citet{2014MNRAS.443.2183F}, (y) \citet{2021ApJ...917...67C}, (z) \citet{2001AJ....121..916T}, (aa) \citet{2009MNRAS.400..168L}, (bb) \citet{2011ApJ...729L..16G}, (cc) \citet{2014ApJ...784..154B}, (dd) \citet{2016AA...587A.109G}, (ee) \citet{2013ApJ...770..145D}, (ff) \citet{2001ApJ...547..323H}, (gg) \citet{2018MNRAS.475..469S}.}
\tablerefs{The following is a list of references for pulsar ephemeris data used in this analysis: CHIME: $\alpha$, Hobart: $\beta$, Jodrell Bank: $\gamma$, MeerKAT: $\delta$, Nancay: $\epsilon$, NICER: $\zeta$, UTMOST: $\eta$, .}
\tablenotetext{a}{The observed $\dot{P}$ has been corrected to account for the relative motion between the pulsar and observer.}
\tablenotetext{b}{The corrected pulsar $\dot{P}$ value is negative, so no value is given and no spin-down limit has been calculated.}
\tablenotetext{c}{This is a globular cluster pulsar for which a proxy period derivative has been derived assuming a characteristic age of $10^9$\,years and a braking index of $n=5$.}
\end{deluxetable}
\end{longrotatetable}

\startlongtable
\movetabledown=2.4cm
\begin{longrotatetable}
\begin{deluxetable}{lrllllllllllll}
\tablecaption{Limits on Gravitational-wave Amplitude, and Other Derived Quantities, for 23 High-value Pulsars from the Three Analysis Methods.\label{tab:highvalue}}
\tabletypesize{\footnotesize}
\tablehead{
\colhead{Pulsar Name} & 
\colhead{$f_{\text{rot}}$} & 
\colhead{$\dot{P}_{\text{rot}}$} & 
\colhead{Distance} & 
\colhead{$h_0^{\text{sd}}$} & 
\colhead{Analysis} & 
\colhead{$C_{21}^{95\%}$} & 
\colhead{$C_{22}^{95\%}$} & 
\colhead{$h_0^{95\%}$} & 
\colhead{$Q_{22}^{95\%}$} & 
\colhead{$\varepsilon^{95\%}$} & 
\colhead{$h_0^{95\%}/h_0^{\text{sd}}$} & 
\colhead{Statistic\tablenotemark{a}} & 
\colhead{Statistic\tablenotemark{b}}\\ 
\colhead{(J2000)} & 
\colhead{(Hz)} & 
\colhead{(s\,s$^{-1})$} & 
\colhead{(kpc)} & 
\colhead{~} & 
\colhead{Method} & 
\colhead{~} & 
\colhead{~} & 
\colhead{~} & 
\colhead{(kg\,m$^2$)} & 
\colhead{~} & 
\colhead{~} & 
\colhead{$^{l=2, m=1,2}$} & 
\colhead{$^{l=2, m=2}$}}
\startdata
\multirow{3}{*}{J0437\textminus4715\tablenotemark{a}\tablenotemark{$\delta$}} & \multirow{3}{*}{173.7} & \multirow{3}{*}{\scinum{1.4}{-20}} & \multirow{3}{*}{0.16 (e)} & \multirow{3}{*}{\scinum{8.0}{-27}} & Bayesian & \scinum{1.3}{-26} & \scinum{3.8}{-27} & \scinum{6.9}{-27} & \scinum{6.6}{29} & \scinum{8.5}{-9} & 0.87 & -4.4 & -2.8 \\ 
 &  &  &  &  & $\mathcal{F}$-statistic & \scinum{2.3}{-26} & \scinum{3.5}{-27} & \scinum{8.9}{-27} & \scinum{8.5}{29} & \scinum{1.1}{-8} & 1.1 & 0.08 & 0.29 \\ 
 &  &  &  &  & 5$n$-vector & \nodata & \nodata & \nodata & \nodata & \nodata & \nodata & \nodata & \nodata \\ 
\hline 
\multirow{3}{*}{J0534+2200\tablenotemark{a}\tablenotemark{$\gamma$}} & \multirow{3}{*}{29.6} & \multirow{3}{*}{\scinum{4.2}{-13}} & \multirow{3}{*}{2.00} & \multirow{3}{*}{\scinum{1.4}{-24}} & Bayesian & \scinum{7.9(6.2)}{-26} & \scinum{6.5(6.1)}{-27} & \scinum{1.3(1.2)}{-26} & \scinum{5.6(5.0)}{32} & \scinum{7.2(6.5)}{-6} & 0.0094(0.0085) & -4.8(-4.9) & -2.6(-2.7) \\ 
 &  &  &  &  & $\mathcal{F}$-statistic & \scinum{7.0(3.1)}{-26} & \scinum{8.8(5.5)}{-27} & \scinum{1.7(1.0)}{-26} & \scinum{7.1(4.2)}{32} & \scinum{9.2(5.4)}{-6} & 0.012(0.007) & 0.65(0.52) & 0.56(0.24) \\ 
 &  &  &  &  & 5$n$-vector & \scinum{1.4(1.3)}{-25} & \nodata & \scinum{9.0(8.7)}{-27} & \scinum{3.7(3.6)}{32} & \scinum{4.9(4.7)}{-6} & 0.0063(0.0061) & 0.02(0.02) & 0.73(0.73) \\ 
\hline 
\multirow{3}{*}{J0711\textminus6830\tablenotemark{a}\tablenotemark{$\delta$}} & \multirow{3}{*}{182.1} & \multirow{3}{*}{\scinum{1.4}{-20}} & \multirow{3}{*}{0.11 (b)} & \multirow{3}{*}{\scinum{1.2}{-26}} & Bayesian & \scinum{1.6}{-26} & \scinum{2.4}{-27} & \scinum{7.0}{-27} & \scinum{4.1}{29} & \scinum{5.3}{-9} & 0.57 & -3.9 & -2.3 \\ 
 &  &  &  &  & $\mathcal{F}$-statistic & \scinum{1.9}{-26} & \scinum{5.2}{-27} & \scinum{9.9}{-27} & \scinum{5.8}{29} & \scinum{7.5}{-9} & 0.81 & 0.17 & 0.34 \\
\enddata
\tablecomments{For references and other notes see Table~\ref{tab:results}. Values in parentheses are those produced using the restricted orientation priors described in Section~\ref{sec:restricted}}
\tablenotetext{a}{For the {\it Bayesian} method this column shows the base-10 logarithm of the Bayesian odds, $\mathcal{O}$, comparing a coherent signal model at both the $l=2$, $m=1,2$ modes to incoherent signal models. For the $\mathcal{F}$-/$\mathcal{G}$-{\it statistic} method this column shows the false-alarm probability for a signal just at the $l=2$, $m=1$ mode, assuming that the $2\mathcal{F}$ value has a $\chi^2$ distribution with 4 degrees-of-freedom and the $2\mathcal{G}$ value has a $\chi^2$ distribution with 2 degrees of freedom. For the 5$n$-{\it vector} method this column shows the $p$-value for a search for a signal at just the $l=2$, $m=1$ mode, where the null hypothesis being tested is that the data are consistent with pure Gaussian noise.}
\tablenotetext{b}{This is the same as in footnote \tablenotemark{a}, but for all the methods the assumed signal model is from the $l=m=2$ mode.}
\tablenotetext{c}{The observed $\dot{P}$ has been corrected to account for the relative motion between the pulsar and observer.}
\tablenotetext{*}{The pulsar J0537\textminus6910 was not analysed in this study but is included for completion. The values listed here are taken from \citet{2021ApJ...913L..27A}.}
\end{deluxetable}
\end{longrotatetable}

\startlongtable
\begin{longrotatetable}
\begin{deluxetable}{lrllll}
\tablecaption{Limits on Gravitational-wave Amplitude from dipole radiation in Brans-Dicke theory for 23  Pulsars using the $\mathcal{D}$-statistic.\label{tab:BDresults}}
\tabletypesize{\footnotesize}
\tablehead{
\colhead{Pulsar Name} & 
\colhead{$f_{\text{rot}}$} & 
\colhead{$\dot{P}_{\text{rot}}$} & 
\colhead{Distance} & 
\colhead{$h_{0d}^{95\%}$} & 
\colhead{FAP}\\ 
\colhead{(J2000)} & 
\colhead{(Hz)} & 
\colhead{(s\,s$^{-1})$} & 
\colhead{(kpc)} &
\colhead{~} &
\colhead{~}}
\startdata
J0437\textminus4715\tablenotemark{a} & 173.7 & \scinum{1.4}{-20} & 0.16 (e) & \scinum{9.7}{-27} & 0.92 \\
 \hline
J0534+2200\tablenotemark{a} &  {29.6} & \scinum{4.2}{-13} & 2.00 & \scinum{9.5(7.0)}{-26} & 0.95(0.31) \\
 \hline
J0711\textminus6830\tablenotemark{a} & 182.1 & \scinum{1.4}{-20} & 0.11 (b) & \scinum{1.9}{-26} & 0.96 \\
 \hline
J0835\textminus4510\tablenotemark{a} & 11.2 & \scinum{1.2}{-13}  & 0.28 (i) & \scinum{1.1(0.74)}{-23} & 0.89(0.39) \\ 
 \hline
J0908\textminus4913 & 9.4 & \scinum{1.5}{-14} & 1.00 (i) & \scinum{1.4}{-22} & 0.89 \\
 \hline
J1101\textminus6101 & 15.9 & \scinum{8.6}{-15} & 7.00 (m) & \scinum{4.7}{-25} & 0.99 \\
 \hline
J1105\textminus6107 & 15.8 & \scinum{1.6}{-14} & 2.36 (b) & \scinum{2.0}{-25} & 0.99 \\
 \hline 
J1302\textminus6350\tablenotemark{a} & 20.9 & \scinum{2.3}{-15} & 2.30 (n) & \scinum{2.3}{-25} & 0.93 \\
\enddata
\tablecomments{For references and other notes see Table~\ref{tab:results}. Values in parentheses are those produced using the restricted orientation priors described in Section~\ref{sec:restricted}. The last column shows the false-alarm probability (FAP) for a signal, assuming that the $2\mathcal{D}$ value has a $\chi^2$ distribution with 2 degrees-of-freedom.}
\end{deluxetable}
\end{longrotatetable}

\allauthors

\end{document}